\def\itc{2}
\def\kavli{3}

\documentclass[onecolumn]{aastex6}

\usepackage{natbib}
\usepackage{graphicx}
\usepackage{graphics}
\usepackage[space]{grffile}
\usepackage{latexsym}
\usepackage{amsfonts,amsmath,amssymb}
\usepackage{url}
\usepackage[utf8]{inputenc}
\usepackage{fancyref}
\usepackage{hyperref}
\usepackage{xcolor}
\usepackage[normalem]{ulem}

\newcommand{\ignore}[1]{}
\begin{document}

\title{One-Armed Spiral Instability in Double-Degenerate Post-Merger Accretion Disks}

%% Notice placement of commas and superscripts and use of &
%% in the author list

\author{Rahul Kashyap\altaffilmark{1}, Robert Fisher\altaffilmark{1,2,3}, Enrique Garc\'{i}a-Berro\altaffilmark{4,5}, Gabriela Aznar-Sigu\'{a}n\altaffilmark{4,5}, Suoqing Ji\altaffilmark{6}, Pablo Lor\'{e}n-Aguilar\altaffilmark{7}}

%\maketitle

\altaffiltext{1} {Department of Physics, University of Massachusetts Dartmouth, 285 Old Westport Road, North Dartmouth, MA. 02740, USA}
\altaffiltext {\itc} {Institute for Theory and Computation, Harvard-Smithsonian Center for Astrophysics, 60 Garden Street, Cambridge, MA 02138, USA}
\altaffiltext {\kavli} {Kavli Institute for Theoretical Physics, Kohn Hall, University of California at Santa Barbara, Santa Barbara, CA 93106, USA}
\altaffiltext{4} {Departament de F\'{i}sica, Universitat Polit\`{e}cnica de Catalunya, c/Esteve Terrades, 5, E-08860 Castelldefels, Spain}
\altaffiltext{5}{Institut d'Estudis Espacials de Catalunya, Ed. Nexus-201, c/Gran Capit\`a 2-4, E-08034 Barcelona, Spain}
\altaffiltext{6}  {Department of Physics, Broida Hall, University of California Santa Barbara, Santa Barbara, CA. 93106-9530, USA}
\altaffiltext{7} {School of Physics, University of Exeter, Stocker Road, Exeter EX4 4QL, UK}

\begin {abstract}
Increasing observational and theoretical evidence points to binary white dwarf mergers as the origin of some if not most normal Type Ia supernovae (SNe Ia). In this paper, we discuss the post-merger evolution of binary white dwarf (WD) mergers, and their relevance to the double-degenerate channel of SNe Ia. We present 3D simulations of carbon-oxygen (C/O) WD binary systems undergoing unstable mass transfer, varying both the total mass and the mass ratio. We demonstrate that these systems generally give rise to a one-armed gravitational spiral instability\ignore{ first presented for the case of a $1.1 + 1.0 M_{\odot}$ C/O binary WD system in an earlier paper \citep{Kashyap2015A}}. The spiral density modes transport mass and angular momentum in the disk even in the absence of a magnetic field, and are most pronounced for secondary-to-primary mass ratios larger than $0.6$. We further analyze carbon burning in these systems to assess the possibility of detonation.  Unlike the case of a $1.1 + 1.0 M_{\odot}$ C/O WD binary, we find that WD binary systems with lower mass and smaller mass ratios do not detonate as SNe Ia up to $\sim8-22$ outer dynamical times. Two additional models do however undergo net heating, and their secular increase in temperature could possibly result in a detonation on timescales longer than those 	considered here.

% Future studies will reveal the complete implication of the long-term fate of these models.

%However, using results from an extended 206 isotope nuclear reaction network, we suggest that most of these systems may ignite carbon if the primary WDs have thin He layer. 
%The condition for detonation of He in this layer yielding a SNe Ia will be possible earlier, during the dynamical phase of the merger.

%and their observational features.   

\end {abstract}


\section {Introduction}
\label{introduction}
Type Ia supernovae (SNe Ia) serve an important role as standardizable cosmological candles \citep{Riess1998A,Perlmutter1999A}\ignore{[REFS?]}. However, we still lack an explanation of their origin and progenitors. Hence, the observed homogeneity of their light curves, which play a key role in the determination of the cosmological parameters, is based on purely empirical grounds. Two pathways are frequently discussed as possible progenitors of SNe Ia. Firstly, a white dwarf (WD) may accrete mass from a main sequence or red giant companion until it reaches the Chandrasekhar mass through the single-degenerate (SD) channel \citep{Whelan_Iben_1973}. Secondly, two WDs may merge and give rise to an SN Ia through the double-degenerate (DD) channel \citep{Iben_Tutukov_1984,Webbink_1984A}.  Contrary to previous expectations, it now appears likely that the SD scenario may account for a wide range of $^{56}$Ni yields, encompassing subluminous through overluminous SNe Ia \citep{Fisher2015A,Childress_2015A}. A variety of observational constraints, including the delay-time distribution (DTD), favors the DD channel \citep{maozbadenes_2010}. Although progress is rapidly being made both observationally and theoretically, there are still many unresolved questions surrounding both channels \citep{Maozetal_2013}. \ignore{[CITE MAOZ ARAA REVIEW.]}  For this reason, other  evolutionary paths that might  produce  a  SN  Ia  outburst have been proposed as alternatives to the SD and the DD scenarios.  Among these  possible channels are the   core-degenerate (CD)  channel  \citep{Sparks_1974,  Livio_2003,  Kashi_2011, Ilkov_2013, Aznar_2015},   and   the  collisional scenario, in which two WDs collide in a dense stellar environment \citep{Raskin2009A,      Rosswog2009A,  Thompson2011A, Kushnir2013A, Aznar_2013A}.
%Due to the inherent stochasticity of ignition within the turbulent convective core of a Chandrasekhar-mass white dwarf, the SD scenario may account for a range of outcomes, from overluminous events such as SN 1991T to subluminous SNe Ia such as SN 1991bg, and even very faint, low-velocity Iax events like SN 2002cx \citep{Fisher2015A}. 
\ignore{[SINCE WE DISCUSS HE IGNITION HERE, ALSO INTRODUCE DOUBLE DETONATION MODEL AT THIS POINT.]} Additionally, the possibility of developing a double detonation in the carbon-oxygen  core of a massive WD through the detonation of a He buffer atop the carbon-oxygen core, and the subsequent shock convergence has also been proposed \citep{Woosley1994A,Livne1990A,Livne1995A}
%, although its robustness in fully three-dimensional numerical simulations has been questioned\citep{Fenn2016A}.

About 10\% of C/O WD binary mergers have a total system mass which exceeds the Chandrasekhar mass \citep{Badenes2012A}. Out of ten such super-Chandrasekhar WD binaries, one will merge within a Hubble time via loss of angular momentum by gravitational waves \citep{Nelemans2001A}. \cite{Toonen_2012A}\ignore{[MORE RECENT REF?]} report super-Chandrasekhar WD binaries to be 1.2\% - 4.3\% of the total number of WD binaries in their binary population models. These super-Chandrasekhar mass binary mergers have been hypothesized to produce a variety of end states -- including SNe Ia \citep{Iben_Tutukov_1984}, accretion-induced collapse (AIC) to neutron stars or anomalous X-ray pulsars \citep{Saio1985A,Miyaji1980A,Rueda_2013A}.
%For $\dot{M}_{acc} \sim 10^{-5}M_{\odot} \rm{yr}^{-1}$, spherically-symmetric models suggest that an off-centered carbon ignition might lead to an AIC and produce a neutron star rather than a SN Ia \citep{Shen2012A,Yoon2007A}. If the mass accretion rate is one order less, compressional heating can cause the temperature to rise at the core and ultimately lead to carbon detonation. 
Some spherically-symmetric models suggested that WD mergers would lead to an off-centered ignition and an AIC \citep{Shen2012A,Yoon2007A}. However, multidimensional simulations reveal a more complex picture of the merger -- namely, a cold WD core surrounded by a disk generated by the tidal disruption of the secondary \citep{Guerrero_2004A,Loren2009A,Pakmor2012A,Raskin2012A,Schwab2012A}, which may possibly lead to a SN Ia. In the most thoroughly-explored scenario, two C/O WDs of nearly equal mass generate sufficient tidal heating to ignite carbon and detonate a sufficiently massive primary \citep{Pakmor_2010,Pakmor_2012,Pakmor2013A}. In the violent merger scenario, a carbon detonation is argued to arise shortly after the tidal disruption of the secondary. The detonation conditions in this case are typically assumed to arise at the location of the maximum temperature in a smoothed particle hydrodynamics (SPH) simulation following the binary through merger, and introduced artificially in a subsequent evolution on an Eulerian grid simulation. However, other authors \citep {raskinetal12} do not find detonation conditions under similar circumstances. Moreover, synthetic light curves and spectra obtained from these violent mergers show a strong dependence upon viewing angle, in tension with observations \citep {molletal14}. 

%The long-term evolution of C/O WD mergers have been explored extensively \citep{Shen2012A}. 

In a previous paper \citep{Kashyap2015A}, we followed the evolution of a $1.1 + 1.0 M_{\odot}$ C/O WD binary beyond the initial stages of evolution resulting from  a  violent merger. In that paper we found that gravitational instability becomes possible in the disk formed by the tidally-disrupted material of the secondary star that surrounds the primary. Using both analytic arguments and numerical simulations, we demonstrated that the disk is particularly susceptible to a one-armed $m = 1$ spiral mode instability. This spiral gravitational instability gives rise to spiral shocks, which carry angular momentum outward, and matter inwards \citep{Blaes_1988}. The accreted matter in turn drives a detonation front through the primary WD. More recent work has similarly found spiral shocks contribute significantly to the angular momentum transport in CVs, even when the magnetorotational instability is active \citep{Ju2016A}.
%We find that for binary white dwarf mergers, such a stability, hereafter referred as an eccentric gravitational stability, is a direct function of mass ratios\citep{Kashyap2015A}. We will present and discuss the results of full non-linear numerical evolution of binary white dwarf mergers in subsequent sections. 

In the linear regime, an axisymmetric perturbation analysis yields the classic criterion that the disk becomes unstable when the Toomre parameter $Q = c_s \kappa / (\pi G \Sigma) < 1$. In this expression, $c_s$ is the sound speed, $\kappa(=\sqrt{d(4\Omega ^2 R) / dR})$ is the epicyclic frequency with $\Omega$ being angular velocity and $R$ is the cylindrical radial distance, $\Sigma$ is the disk surface density, and $G$ is the gravitational constant.  Previous numerical and analytic studies have extended the classic axisymmetric Toomre condition to non-axisymmetric perturbations. For instance, in a pioneering study, \cite{Adams_1989} and \cite{Shu_1990} demonstrated that the criterion for the instability to develop the $m=1$ spiral mode is that the Toomre parameter $Q$ should be smaller than 3 at corotation. Two key questions to be addressed in this paper are the following ones. To what extent is the development of the eccentric one-armed spiral instability a general outcome of binary WD mergers? Under what circumstances may this instability lead to detonation of the C/O WD core? In our previous paper, we argued from general principles that the inner portion of the WD disk is marginally susceptible to gravitational instability over a wide range of mass ratios, and completely independent of the total system mass. However, while we conducted  a careful set of simulations varying the numerical resolution and timestep criteria in \cite{Kashyap2015A}, we studied just one WD merger model  of a $1.1 + 1.0 M_{\odot}$ C/O WD binary. In this paper, we present the hydrodynamical evolution of several C/O WD mergers with different masses and mass ratios over several outer dynamical timescales, with the goal of  analyzing the development of the spiral mode instability in these systems, and investigating their prospects to produce SN Ia.

The plan of the paper is as follows. In \S \ref{num_sim}, we present our numerical methodology and suite of binary WD models. In \S \ref{results}, we present and discuss the results of the full nonlinear numerical evolution of these binary WD mergers.
% and we discuss the ignition conditions obtained from an extended nuclear reaction network including $\alpha$-capture. 
We then consider the possibility of carbon ignition with the set of initial conditions of the models considered. Lastly, in \S \ref{conclusion}, we summarize our findings and elaborate our conclusions.

\section{Methodology}
\label{num_sim}
We conducted a suite of simulations of merging C/O WDs, which initially had equal abundances of carbon and oxygen, and with masses  $1.1 + 1.0 M_{\odot}$, $1.0 + 0.9 M_{\odot}$, $0.8 + 0.6 M_{\odot}$, $0.8 + 0.5 M_{\odot}$ and $1.0 + 0.6 M_{\odot}$. %Models have SPH resolution of $2\times 10^5$ particles.
For these simulations, we used the SPH code employed in our previous studies \citep{Loren_Aguilar_2010}, with a resolution of $2\times 10^5$ SPH particles. The SPH simulations utilizes a nuclear $\alpha$-network which incorporates 14  nuclei: He, C,
O, Ne,  Mg, Si,  S, Ar,  Ca, Ti, Cr,  Fe, Ni,  and Zn. The reactions
considered are $\alpha$ captures and their associated back
reactions, C-C and C-O fusion reactions, and a quasi-equilibrium reduced $\alpha$-network for temperatures higher than $3 \times 10^9$K \citep{Hix_1998A}. All reaction rates are taken from
the \texttt{REACLIB} database \citep{Cyburt_2010}. Neutrino losses are also taken into account, using \citet{Itoh_1996}.\\
The initial SPH conditions to simulate the merger of the two white dwarfs in a synchronized binary orbit closely follow the procedure outlined in \cite{Dan2011A}. In particular, we first set the coalescing white dwarfs in the co-rotating frame at a sufficiently large distance, to avoid premature mass transfer. Afterwards, the orbital separation is slowly decreased in such a way that the orbital shrinkage time is always substantially longer than the dynamical timescale of the secondary. As soon as the first SPH particle of the secondary white dwarf reaches the inner Lagrangian point (or, equivalently, the secondary fills its Roche lobe) the relaxation process is stopped, and the simulation of the merger starts. In this way we obtain accurate initial separations for our simulations. The distances between the centers of mass of the merging white dwarfs when mass transfer begins, as well as some other details,  are listed in Table 1.\\ 
We introduce a lower limit for the density ($\rho_{\min} = 6 \times 10^3$ g cm$^-3$), corresponding to a maximum smoothing length of $\sim 1400$ km. This density limit was introduced into our SPH code in order to prevent large velocity dispersions in particles being ejected during the merging process, and a consequent greatly-reduced timestep. This maximum smoothing length is in a sense analogous to the base level grid of an adaptive mesh refinement code, which imposes a coarsest possible mesh size to a grid-based calculation. The amount of matter affected by the density limit is extremely small when compared with the mass present in the main body of the merger, so its impact on the development of the spiral instability is minimal. However, one must be cautious when interpreting the outermost regions of each disk, since the density profile inevitably becomes underresolved at some radius.

Our SPH simulation results are in good overall agreement with other similar models up to the point of merger \citep{Loren2009A} although our peak temperatures are somewhat higher than those of \cite{Dan2011A}. Most previous SPH calculations of the coalescence of white dwarfs obtain the temperature following the evolution of the internal energy. From the newly-computed internal energy, the temperature can be obtained by inverting the equation of state. However, since a large portion of the merging white dwarfs is degenerate, this procedure can produce poor results in some cases. To avoid such spurious results we follow the evolution of the temperature simultaneously using the specific heat -- see Eqs. (5) and (6) of \cite{Aznar_2013A}. Specifically, we always evolve the temperature using both prescriptions, and if the difference between the temperatures is larger than 5\%, and the degeneracy parameter is sufficiently large so that the electrons can be considered as partially degenerate, we consider that the temperature determination obtained employing the specific heat is more reliable, and we adopt this value. Otherwise, if this is not the case, we compute the temperature in the standard way:  that is, inverting the equation of state. Using this dual energy description, total energy is best conserved, and the numerical results are more reliable than an internal energy description.
 
\begin{table}[h!]
\renewcommand{\thetable}{\arabic{table}}
\centering
\caption{Runs in this paper. The maximum finest resolution for each model is 136~km. The maximum temperature, $T_{\rm max}$, and the density at maximum temperature at the end of the simulation, $\rho$ at $T_{\rm max}$, are also listed. For the detonating model, $1.1 + 1.0 M_{\sun}$, the data is shown at a time $t = 108.98$ s just prior to detonation. }
\vspace{0.5cm}
\resizebox{\linewidth}{!}{
\begin{tabular}{cccccc}
\tablewidth{0pt}
\hline
\hline
Primary Mass  & Secondary Mass & Mass ratio & $ T_{\rm max}$ &  $\rho$ at  $T_{\rm max}$ & Initial Distance  \\ 
($M_{\sun}$)  & ($M_{\sun}$) &  & ($\times 10^9 $K) & ($\times 10^6 $~gm/cm$^3$)  & $10^{-2}$ R$_{\odot}$\\
\hline
1.1 & 1.0 & 0.91 & 3.2 & 6.7 & 2.327 \\
1.0 & 0.9 & 0.90 & 1.0 & 3.3 & 2.594 \\
0.8 & 0.6 & 0.75 & 1.3 & 0.7 & 2.561 \\
0.8 & 0.5 & 0.63 & 0.4 & 0.7 & 3.067 \\
1.0 & 0.6 & 0.60 & 0.7 & 1.4 & 2.761 \\
\hline
\end{tabular}}
\label{table:runtable}
\end{table}

%%------------ TABLES------------------%%

%In this paper, we define $t=0$ to be the time of maximum temperature in the SPH simulation.  
We subsequently mapped the SPH data at about 1.5 outer disk dynamical times, or 40~s after merger, onto a 3D Eulerian grid using the adaptive mesh refinement (AMR) code FLASH 4.1 \citep{Fryxell2000A,Calder2002A, Dubey_2009A}. This time of remapping defines $t = 0$ in this paper. Our domain extends from $-2.8\times 10^{10}$ cm to $+2.8\times 10^{10}$~cm in each direction. At the time of remapping, the maximum temperature is typically $1.2 \times 10^9$~K. We refine spatial regions in which any cell has a mass exceeding $4\times 10^{27}$~g. Both the SPH and FLASH models employ an identical equation of state which includes contributions from electrons, nuclei, and photons, and considers an arbitrary degree of special relativistic and degeneracy effects for the electrons \citep{Timmes2000A}. The use of the same EOS in both cases avoids spurious (and undesirable) effects introduced by the remapping procedure. The FLASH models include nuclear energy generation using a 19 isotope $\alpha$-chain reaction network \citep{Timmes1999A}.
%\footnote {Later, in \S \ref{ignition_condn} we will employ analytical estimate when assessing C ignition.} 
FLASH incorporates both multigrid and multipole solvers to model self-gravity. In this paper, we use an improved multipole solver \citep{Couch_2013} with isolated boundary conditions and including terms through $\ell = 60$ in the multipole expansion for our FLASH simulations. In \cite{Kashyap2015A}, we found excellent agreement between the accuracy of the multipole and multigrid solvers for this application, which justifies the selection of the multipole solver in this paper. Fluid boundary conditions follow a diode boundary condition, which assumes zero hydrodynamic gradients across the outer domain boundaries while also enforcing zero inwardly-directed hydrodynamic fluxes across the domain boundary. 

The early phases of the merger are challenging to accurately treat with mesh-based codes, which experience larger advection errors in the presence of strong density gradients than SPH codes \citep {taskeretal08}, and suffer additional artifacts such as spurious torques aligning disks with the Cartesian axes \citep {hahnetal10}. Additionally, fluid instabilities and  shocks are notoriously difficult to accurately simulate using SPH techniques, possibly leading to non-convergence in some instances \citep {springel10}.  Our approach to simulating WD mergers therefore combines the two approaches in the regimes in which they are best suited: SPH during the early phases of the merger up to the formation of the tidally-disrupted disk, and AMR in the later stages subsequent to disk formation, in which accurate treatment of the disk spiral instabilities are crucial. The transition is made at peak compression in the SPH simulation.

We follow the system evolution for 430~s  subsequent to merger, which corresponds to $\sim$ 8 -- 22 outer dynamical times for the models presented here. Detailed results of the $1.1 M_{\sun} + 1.0 M_{\sun}$ C/O WD binary have been presented in our previous publication \citep{Kashyap2015A}. Here, we compare this model with other models of lower system mass and mass ratios. In particular, we analyze the time evolution of midplane density and temperature structure of all models. We present the results of these analyses in the next section.

\section{Results}
\label{results}

The eccentric instability is driven by the gravitational tug of the indirect potential of the primary by the disk produced by the tidally-disrupted secondary \citep{Adams_1989}. Based on theoretical considerations, we expect the gravitational instability to decrease as the secondary WD mass becomes significantly smaller than that of the primary \citep{Kashyap2015A}. We now consider how this expectation is borne out using fully multidimensional simulations of a suite of binary WD mergers, and explore the development of the instability as a function of the system mass, $M$, and of the mass ratio, $q$.

In \S \ref{midplane}, we analyze the growth of density perturbations by taking density slices in the midplane. To quantify the growth of the instability, we plot the Fourier modes as a function of time for different mass ratios. We consider the stability of the disks in our model at $t=0$ using Toomre's $Q$ parameter, which was defined previously in \S \ref{introduction}. In \S \ref{heating}, we then present the evolution of the nuclear energy output and of the maximum temperature as the result of the spiral instability and resultant angular momentum transport and mass accretion, in the disk. \ignore{We then study the maximum temperature, $T_{\rm max}$, and the density at maximum temperature, $\rho_{\rm max}$, for all the simulations presented here.} In \S \ref{ignition_condn}, we employ an analytic criterion for carbon ignition to assess the prospects for detonation of the primary WD.

%, obtained employing analytical estimates widely used in the literature.

\subsection{Midplane Structure and Global Quantities}
\label{midplane}

\begin{figure}
	\begin{center}
		\includegraphics[width=0.6\columnwidth]{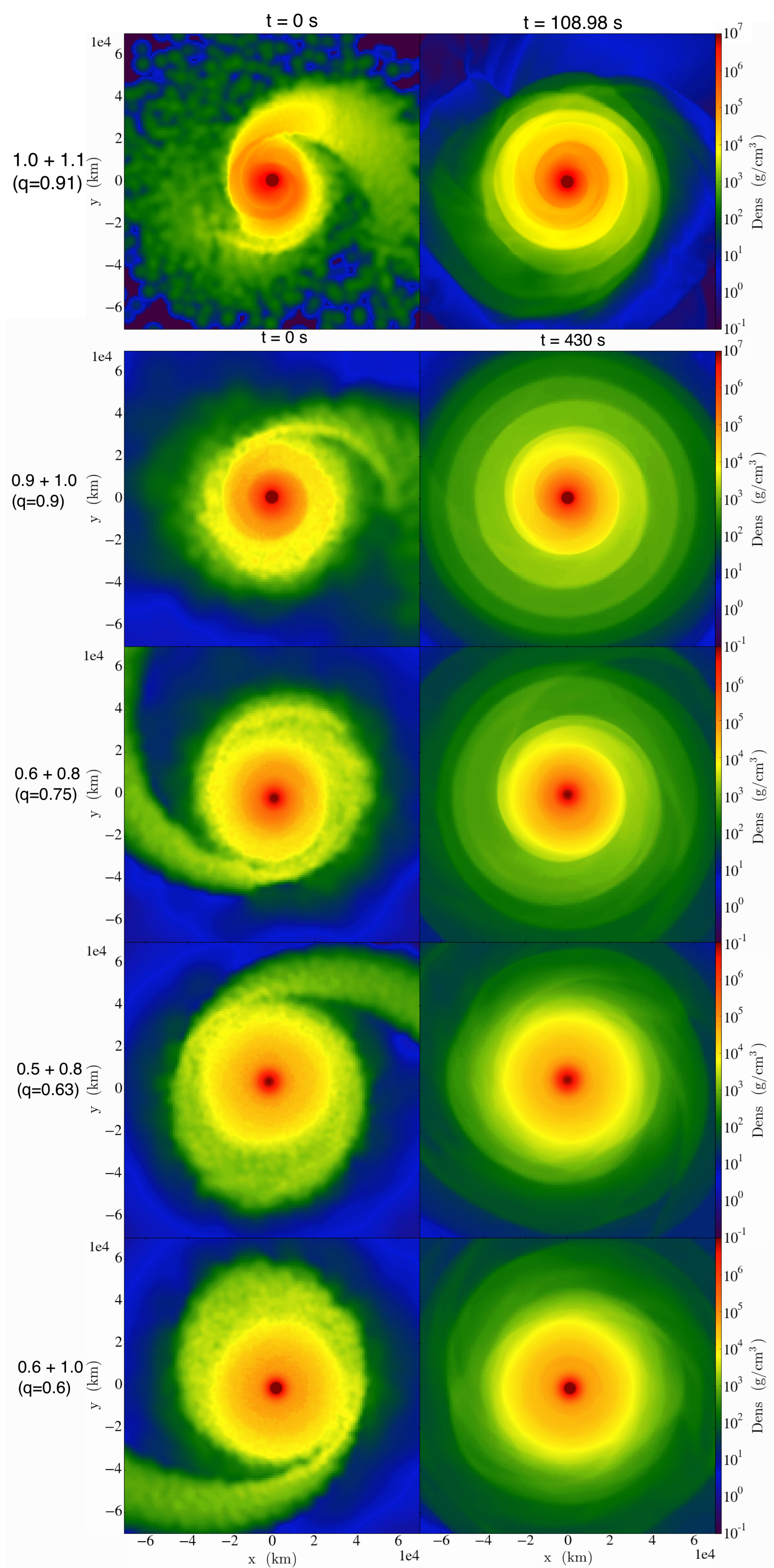}
		\caption{A detailed view of the disk midplane density at $t=0$ and $t=430$s for different mass combination and mass-ratios. Models presented from top to bottom are -- $0.9 M_{\odot}$ + $1.0 M_{\odot}$, $0.6 M_{\odot}$ + $0.8 M_{\odot}$, $0.5 M_{\odot}$ + $0.8 M_{\odot}$ and $0.6 M_{\odot}$ + $1.0 M_{\odot}$. These plots show the increasing strength of the spiral disk instability with mass ratio.}
		\label{fig:allMass_disk}
	\end{center}
\end{figure}

We observe the development of spiral structures in the disk caused by the eccentric spiral mode gravitational instability. Figure~\ref{fig:allMass_disk} shows the initial and the final (at $t=430$~s) density structure at the disk midplane for the four new models considered here. Models are presented for decreasing mass ratios, from top to bottom. Models with higher mass ratios produce  more tightly wound spirals than lower mass ratios.
%, and their development does not sensitively depend upon the total system mass. 
Overall, the structure is symmetric with respect to the $z=0$ plane.

In \cite{Kashyap2015A}, we found that the spiral modes cause matter to be accreted onto the primary WD and angular momentum to be transported outward. We quantify the growth of the spiral structure by decomposing the Fourier components of the surface disk density into the lowest-order azimuthal modes. The disk surface density $\Sigma (r, \phi)$ is the vertically-integrated mass density $\rho (r, \phi, z)$ over scale height, $H$ :
\begin {equation}
\Sigma (r, \phi) = \int_{-H}^H \rho (r, \phi, z)\, dz 
\end {equation}
We then define the Fourier coefficients $C_m$ of the surface density over azimuthal angle:
\begin {equation}
C_m = \frac{1}{2\pi} \int_{0}^{2\pi} e^{im\phi} \, d\phi \int_{R_{\rm min}}^{R_{\rm max}}  \Sigma(r,\phi) r \, dr  
\end {equation}
The normalization is chosen so that $C_0$ simply represents the total mass of the disk. Here $R_{\rm min}$ is taken to be tidal radius of respective models and $R_{\rm max}$ is the domain boundary. The tidal radius is defined to be the distance at which the tidal force of the primary star overcomes the self-gravity force of the secondary, and is given by $R_{\rm tid}=q^{1/3}R_{2}$ where $q$ is the mass-ratio and $R_{2}$ is the radius of the secondary WD.  The higher-order coefficients represent the amplitude of each mode. We plot the coefficients, normalized to the disk mass, $C_m / C_0$, as a function of time $t$ for the first four modes $m = 1$  to 4 (Figure~\ref{fig:mode_amp}). The evolution of the coefficients in time depends sensitively upon both, total mass and mass ratios. However, the strength of the time averaged $m=1$, 2, 3 and 4 modes is found, in general, to increase with mass ratio (see Table~\ref{table:mode_amp}). 

Semi-periodic oscillations in the amplitudes of the $m = 1$ and higher-order modes are clearly seen in figure~\ref{fig:mode_amp}, with the frequency of the oscillations increasing with the inner orbital frequency of the disk. Semi-periodic oscillations in spiral mode amplitudes have been noted previously in the literature e.g. figure 17 of \cite{Nelson1998}. The time-variability of the $m = 1$ mode amplitude has been interpreted as arising due to the intrinsically non-stationary nature of mode-mode interactions with higher-order modes within the disk. \cite{Laughlin1997} modeled the mode-mode interactions using analytic governing equations, and confirmed the time evolution obtained analytically using fully hydrodynamic models. As a consequence, the non-linear development growth and saturation of spiral modes as seen here generally lead to a more complex, non-steady time evolution than is captured in simplified analytic linear growth models.  

 In Figure~\ref{fig:disk_prop_all}, the Toomre $Q$ parameter is plotted for all models extending in the disk region, starting from their respective tidal radii.  The growth of the modes depends on the fraction of the disk that is below $Q=3.0$, which is largest in the case of a $1.0 M_{\sun}$ + $1.1 M_{\sun}$ binary system consistent with the theoretical model. For systems with lower mass ratios, smaller portions of the disk are unstable by this criterion. Thus, the nonlinear evolution of these systems leads to the development of a spiral-mode instability in the resulting disks, as predicted from their initial properties. %Mass accretion happens slowly in the form of a stream which forces the mixing of hot disk material into the cold degenerate matter of white dwarf. 

\subsubsection{Numerical Issues Regarding Disk Dynamics}
As mentioned in the methodology section, simulating disks in a Cartesian geometry on an Eulerian mesh can lead to a number of numerical issues.
These numerical issues become particularly important when studying non-axisymmetric modes for many outer-disk orbital times, as we do here. For instance, \cite{Krumholz2004} found that spurious advection in a disk simulated in Cartesian geometry leads to a number of numerical artifacts, including the numerically-driven viscous spreading of a thin ring, and the rapid evacuation of the innermost zones of a disk. Additionally, angular momentum conservation is generally an issue with a simulation on an Cartesian geometry Eulerian mesh \citep {trueloveetal98}.  Here, we describe how we  have addressed these concerns by building upon and extending our convergence tests published in \cite{Kashyap2015A} by exploring the behavior of the $m = 1$ mode in one of our simulations, consisting of the $1.1 M_{\odot} + 1.0 M_{\odot}$ WD merger.

The numerical and artificial viscosities in modern hydrodynamic codes like FLASH are nonlinear and generally unknown functions of resolution and the state of the gas. However, a key property which both numerical and artificial viscosity share is that they both must be functions of the resolution. Specifically, if the angular momentum transport leading to the $m = 1$ mode generation seen here is numerically-driven, the angular momentum transport will be a strong and generally monotonic function of spatial resolution. That is, if the $m = 1$ mode is artificial, as the mesh is coarsened, the $m = 1$ mode will generally increase in amplitude. Further, if the $m = 1$ mode is artificial, then as the mesh spacing goes to zero, so too will the amplitude of the $m = 1$ mode. Conversely, if the results are physical, then as the mesh spacing goes to zero, the amplitude of the $m = 1$ mode will converge to a single value at a given time.\\

\begin{figure}
	\begin{center}
		\includegraphics[width=0.8\columnwidth]{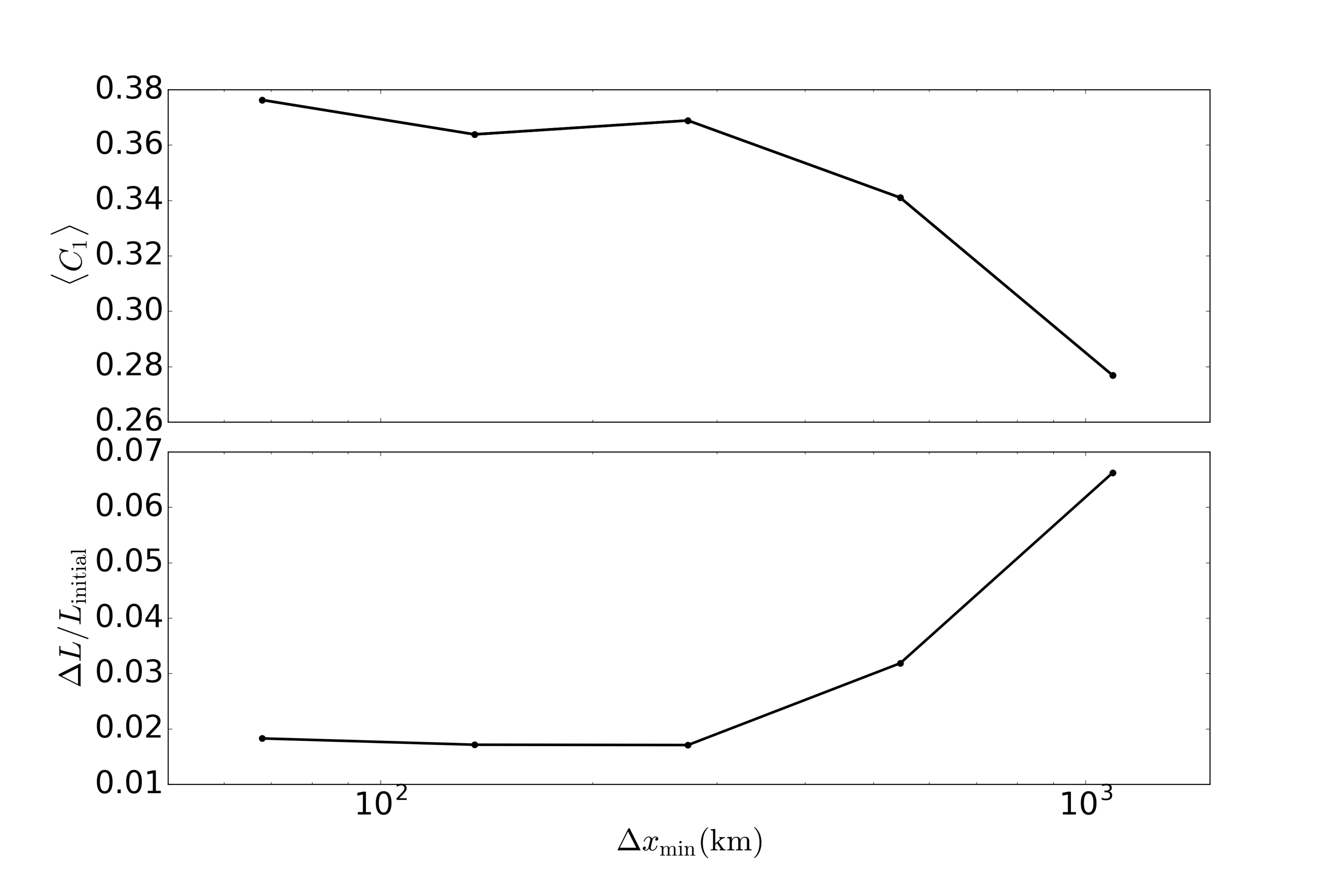}
		\caption{Convergence studies for the 1.0 M$_\odot$ + 1.1 M$_{\odot}$ model. The top panel shows the time-averaged RMS coefficient for $m=1$ mode for different resolution runs. The bottom panel shows angular momentum deviation as compared to initial angular momentum for different resolutions. All runs have been performed at the fiducial resolution of 136 km for the time interval leading up to detonation for the finest resolution run.}
		\label{fig:convergence}
	\end{center}
\end{figure}

We first address to what extent the angular momentum is conserved over the duration of the simulations considered. We have monitored the angular momentum in the course of evolution to test spatial convergence, and find that deviation of angular momentum conservation from start to finish is always better than $\sim 10\%$, and is typically at the $\sim1\%$ level for the finest resolutions considered. A convergence study of the angular momentum conservation is plotted for the $1.1 M_{\odot} + 1.0 M_{\odot}$ model in figure \ref{fig:convergence}. Note that this plot shows the angular momentum error as a function of finest grid resolution, with the finest resolution models approaching $\sim 2\%$ error. The reason why this error does not asymptote to zero is due to the construction of the resolution study, which keeps the coarser grids at the same level of resolution while only refining the center-most regions.\\

We next examine the convergence behavior of the $m = 1$ mode. In figure \ref{fig:convergence}, we plot the time-averaged RMS amplitude of the $m = 1$ mode over four doublings in maximum spatial resolution, while maintaining the same coarse mesh structure.  It is evident that as the resolution is increased, so too is the amplitude of the $m = 1$ mode, which is opposite of the behavior one would expect from angular momentum transport driven by numerical artifacts. Additionally, the time-averaged RMS amplitude converges to within $\sim 5\%$ at the highest resolutions considered. This convergence study supports the conclusion that our findings are driven by disk physics and not numerics.

\subsection{Heating and Nuclear Burning}
\label{heating}

\begin{figure}
	\begin{center}
		\includegraphics[width=1.0\columnwidth]{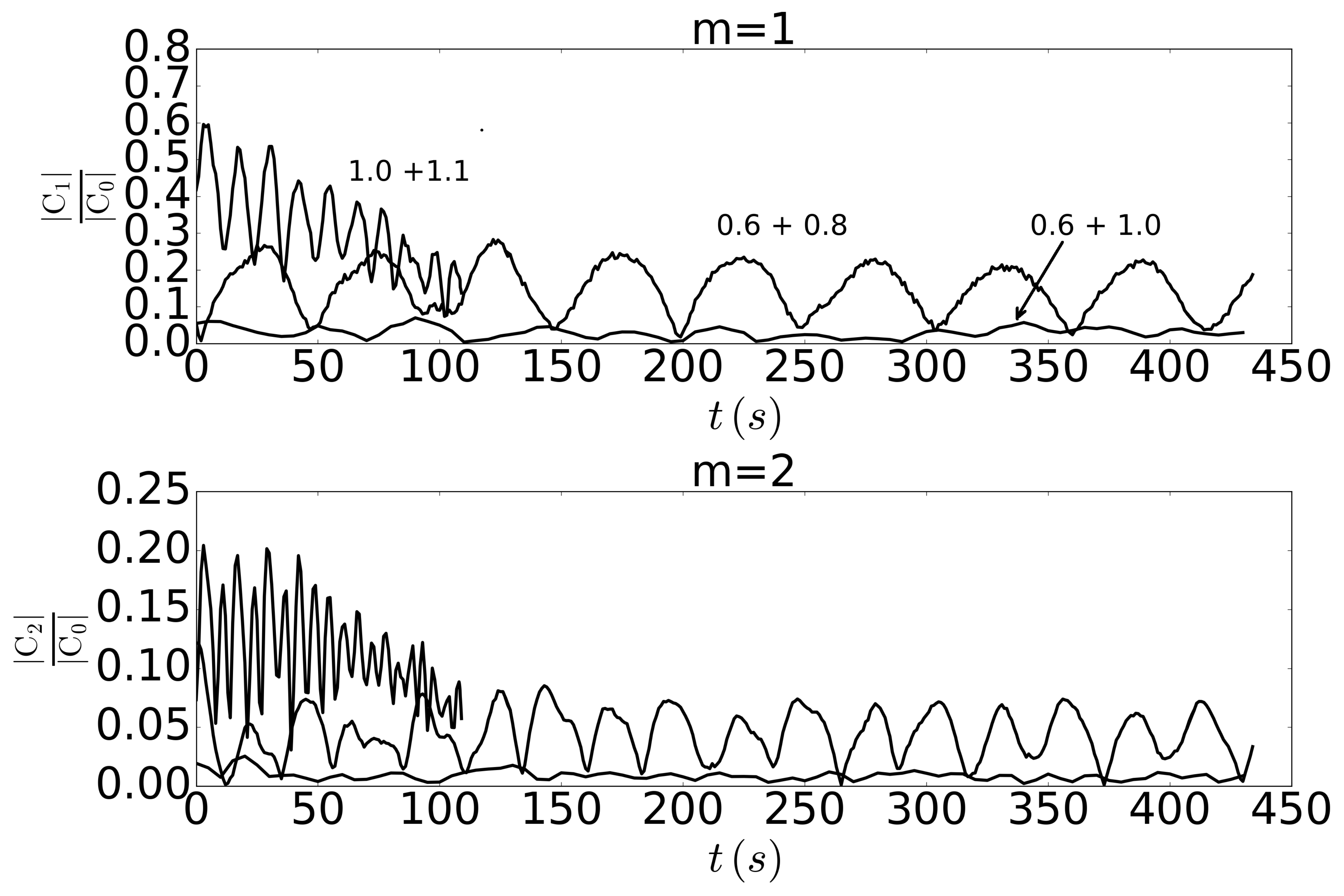}
		\caption{The growth of lowest two non-axisymmetric modes in the disk shown for three representative models. Their time-integrated values are presented in Table \ref{table:mode_amp} . }
		%(legend for this and all subsequent figures -- $1.0 M_{\odot}$ + $1.1 M_{\odot}$  magenta, $0.9 M_{\odot}$ + $1.0 M_{\odot}$ cyan, $0.6 M_{\odot}$ + $0.8 M_{\odot}$  red, $0.5 M_{\odot}$ + $0.8 M_{\odot}$ green and $0.6 M_{\odot}$ + $1.0 M_{\odot}$ blue). }
		\label{fig:mode_amp} 
	\end{center}
\end{figure}

\begin{figure}
	\begin{center}
		\includegraphics[width=0.9\columnwidth]{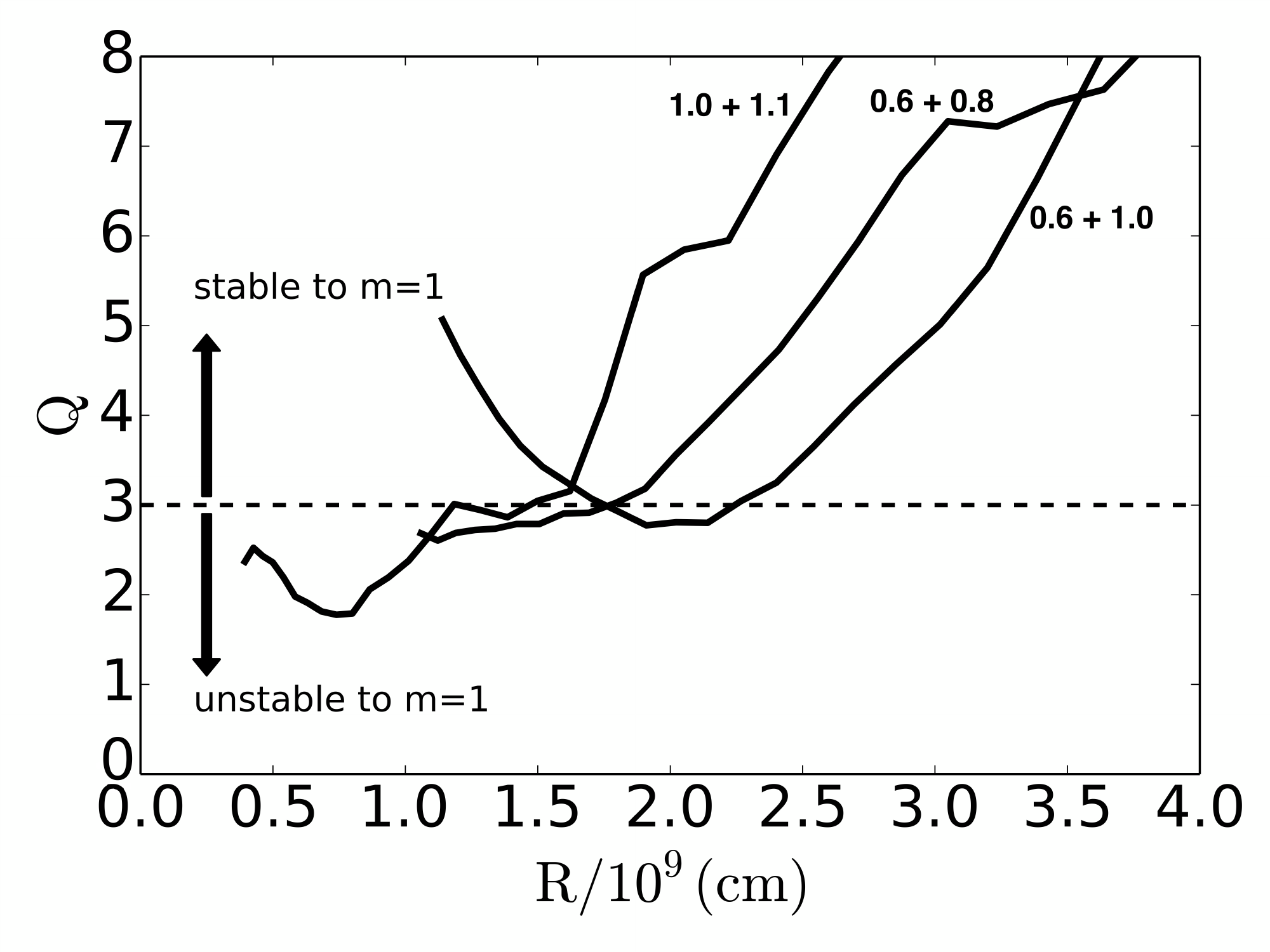}
		\caption{The Toomre $Q$ value, as defined in the text, in the disk region for three representative models. The analytic estimate for $m = 1$ instability ($Q \textless  3$ at corotation) is indicated. For all models, the curve starts at their respective tidal radius calculated as $R_{\rm tid}=q^{1/3}R_{2}$, where $R_{2}$ is the secondary radius. The $1.0 M_{\odot}$ + $1.1 M_{\odot}$ model has the largest portion of its disk unstable by this criterion.} 
		\label{fig:disk_prop_all}
	\end{center}
\end{figure}

Figure~\ref{fig:temp_disk} shows the midplane temperature slices at two points in the evolution, $t=0$ and $t=430$~s for all models, apart from the detonating $1.1 \,M_{\sun}$ + $1.0\, M_{\sun}$ model, which is shown at the onset of detonation at $t = 108.98$ s. As can be seen in these plots, the tidally-heated disk surrounds the cold primary, which remains almost intact. The maximum temperature always occurs just outside the surface of the primary star, consistent with a virial estimate \citep {Kashyap2015A}. Typically the maximum temperature is $\sim$ $10^{8}$~K at $t=0$. The typical density at the location of maximum temperature is $\sim$ $10^{5}$~g~cm$^{-3}$. The temporal evolution of the maximum temperature is shown in Fig. ~\ref{fig:maxtemp_evol} for all models. The maximum temperature increases with increasing total mass -- namely, the configurations resulting from the mergers of the $1.1 \,M_{\sun}$ + $1.0\, M_{\sun}$, $0.9 \, M_{\sun}$ + $1.0 \, M_{\sun}$ and $0.5\, M_{\sun}$ + $0.8\, M_{\sun}$ systems. However, only the first of these models shows a prominent increase of the maximum temperature as a consequence of mass accretion due to the spiral instability. This effect is seen most clearly in the disk region in the first slice. The rest of the systems show modest temperature increases. Finally, the models with the smallest total masses do not experience any marked increase of the maximum temperature. The increase of temperature in the C/O WD surface, as hot disk material is accreted, leads for the first of our models to the conditions necessary for C to detonate  -- $T \sim 2 \times 10^9$~K at densities $\rho \sim 10^7$~g~cm$^{-3}$ \citep {Kashyap2015A,Seitenzahl_2009}. We find that a detonation arises only in the systems with large mass ratios that have higher maximum temperature and density at $t=0$~\citep{Kashyap2015A}. However unlike previous results \citep{Dan2014A}, we find an increase of maximum temperature which in this case can be attributed to the spiral mode instability, whose saturated amplitude depends sensitively  upon the mass ratio (see Table~\ref{table:mode_amp} and Fig.~\ref{fig:mode_amp}). 

We emphasize that systems with the same primary masses but differing mass ratios evolve distinctly under the action of the spiral instability. For instance, consider the temperature slices in the first and last rows of Fig.~\ref{fig:temp_disk} (corresponding to the $1.0\, M_{\sun} + 0.9\, M_{\sun}$ and $1.0\, M_{\sun} + 0.6\, M_{\sun}$ simulations). They have the same primary masses but, owing to their different mass ratios, they exhibit different behaviors. For example, the mode strength and temperature in the disk region is larger for the $1.0\, M_{\sun} + 0.9\, M_{\sun}$ case. There is a larger increase of the maximum temperature for the $1.0\, M_{\sun} + 0.9\, M_{\sun}$ case when compared to the $1.0\, M_{\sun} + 0.6\, M_{\sun}$ model. Similarly, we find that the growth rate of the instability for the $0.8\, M_{\sun} + 0.6\, M_{\sun}$ model is greater than that of the $0.8\, M_{\sun} + 0.5 M_{\sun}$ system.

%Choice of the final time is motivated by the evolution of disk system for about $\sim10$ dynamical time. 
To explain these results we integrated the nuclear energy released during the evolution and computed the cumulative nuclear energy yield as a function of time. The evolution of the cumulative energy release, including neutrino losses, is shown in Fig.~\ref{fig:energy_evol}. Of all the models considered here, we find that only the cases of $1.0\, M_{\sun} + 1.1\, M_{\sun}$ and $0.9\, M_{\sun} + 1.0\, M_{\sun}$ have a positive energy balance at all times during its evolution. The model resulting from the merger of a $0.6\, M_{\sun} + 1.0\, M_{\sun}$ binary system has positive energy output at later times. We analyzed the time evolution of mass and internal energy fluxes in different cylindrical radial bins for the 1.0+0.6 model. In this model, heating from the compression in the inner disk, which is concentrated within the spiral shocks, advects hot material inwards towards the primary and is responsible for the slow secular increase in temperature, while the other cases, $0.6\, M_{\sun} + 0.8\, M_{\sun}$ and $0.5\, M_{\sun} + 0.8\, M_{\sun}$ have negative energy balances (due to neutrino cooling) and hence show slight decreases of the total internal energies and maximum temperatures (see Fig.~\ref{fig:maxtemp_evol}).
%%%------------ TABLES------------------%%
%\begin{deluxetable}{ccccccc}
%\tablecolumns{7}
%\tablewidth{0pt}
%\tablecaption{Time Averaged Normalised Amplitudes for Different Models. }
%\tablenum{2}
%\tablehead{\colhead{Primary Mass} & \colhead{Secondary Mass} & \colhead{Mass Ratio} & \colhead{$m=1$} & \colhead{$m=2$} & \colhead{$m=3$} & \colhead{$m=4$} \\ 
%\colhead{($M_{\odot}$)} & \colhead{($M_{\odot}$)} & \colhead{} & \colhead{($\frac{|C_{1}|}{|C_{0}|}$)} & \colhead{($\frac{|C_{2}|}{|C_{0}|}$)} & \colhead{($\frac{|C_{3}|}{|C_{0}|}$)} & \colhead{($\frac{|C_{4}|}{|C_{0}|}$)}}%\begin{tabular}{|c|c|c|c|c|c|}
%\startdata
%
%1.1 & 1.0 & 0.91 & 0.4694 & 0.1865 & 0.0674 & 0.0689\\
%1.0 & 0.9 & 0.90 & 0.1593 & 0.0570 & 0.0209 & 0.0082\\
%0.8 & 0.6 & 0.75 & 0.1650 & 0.0494 & 0.0171 & 0.0073\\
%0.8 & 0.5 & 0.63 & 0.0440 & 0.0121 & 0.0060 & 0.0042\\
%1.0 & 0.6 & 0.60 & 0.0331 & 0.0097 & 0.0062 & 0.0042\\
%
%\enddata
%%\end{tabular}
%\label{table:mode_amp}
%\end{deluxetable}%
%%%------------ TABLES------------------%%
\begin{table}[t]
\centering
\caption{Time averaged normalized amplitudes for all the different models presented in this paper. }
\vspace{0.5cm}
\resizebox{1\linewidth}{!}{
\begin{tabular}{ccccccc}
\hline
\hline
%\tablenum{1}
Primary Mass & Secondary Mass & Mass ratio & $m=1$ & $m=2$ & $m=3$ & $m=4$ \\ 
($M_{\sun}$) & ($M_{\sun}$) &  & $\left|\frac{C_{1}}{C_{0}}\right|$ & $\left|\frac{C_{2}}{C_{0}}\right|$ & $\left|\frac{C_{3}}{C_{0}}\right|$ & $\left|\frac{C_{4}}{C_{0}}\right|$\\
\hline
%\hline
1.1 & 1.0 & 0.91 & 0.4694 & 0.1865 & 0.0674 & 0.0689\\
1.0 & 0.9 & 0.90 & 0.1593 & 0.0570 & 0.0209 & 0.0082\\
0.8 & 0.6 & 0.75 & 0.1650 & 0.0494 & 0.0171 & 0.0073\\
0.8 & 0.5 & 0.63 & 0.0440 & 0.0121 & 0.0060 & 0.0042\\
1.0 & 0.6 & 0.60 & 0.0331 & 0.0097 & 0.0062 & 0.0042\\
\hline
\end{tabular}}
\label{table:mode_amp}
\end{table}
%%------------ TABLES------------------%%

\subsection{Ignition} 
\label{ignition_condn}

The slow secular increase of the maximum temperature observed in some of our merger models poses a key question: Under what conditions will carbon ignite, and lead to a subsequent detonation of the underlying CO WD core?  Carbon requires high temperatures ($> 2 \times 10^9$~K at densities $> 10^7$~g~cm$^{-3}$) to ignite, which are very difficult to achieve naturally in sub-Chandrasekhar models, including double-degenerate mergers. Indeed, this difficulty in reaching carbon ignition conditions has been a primary motivation in pursuing the violent merger scenario itself, and in pursuing alternatives to violent mergers, including head-on WD collisions \citep{Raskin2009A,Pakmor_2010,Pakmor_2012,Kushnir2013A}, and tidal  disruption of WDs around intermediate-mass black holes \citep{Guillochon_2015A}. 
%The spiral instability drives an accretion flow of hot material, which in turn leads to unstable carbon burning at sufficiently high densities to ignite detonations in WD binary systems with high mass ratios.
The Zel'dovich mechanism, which invokes the development of a subsonic burning front as it accelerates into a detonation front across a temperature gradient, is the primary mechanism for the initiation of a detonation front in an unconfined medium \citep{Zeldovich_1970A}. A full analysis yields the critical length scale as a function of density and temperature, with a somewhat weaker dependence on the functional form of the temperature profile, e.g. linear or Gaussian \citep{Seitenzahl_2009}. At low densities and temperatures, the critical length scale is comparable to the size of the star itself, making any such detonation implausible. However, because of the extreme sensitivity of the nuclear burning rates to temperature, at sufficiently high temperatures,  the critical length scale drops by many orders of magnitude. Consequently, when the critical length scale of the detonation becomes smaller than the hydrodynamic scales of interest, any small portion of the flow with a well-ordered temperature gradient is likely to form a detonation front. 
\begin{figure}
	\begin{center}
		\includegraphics[width=0.6\columnwidth]{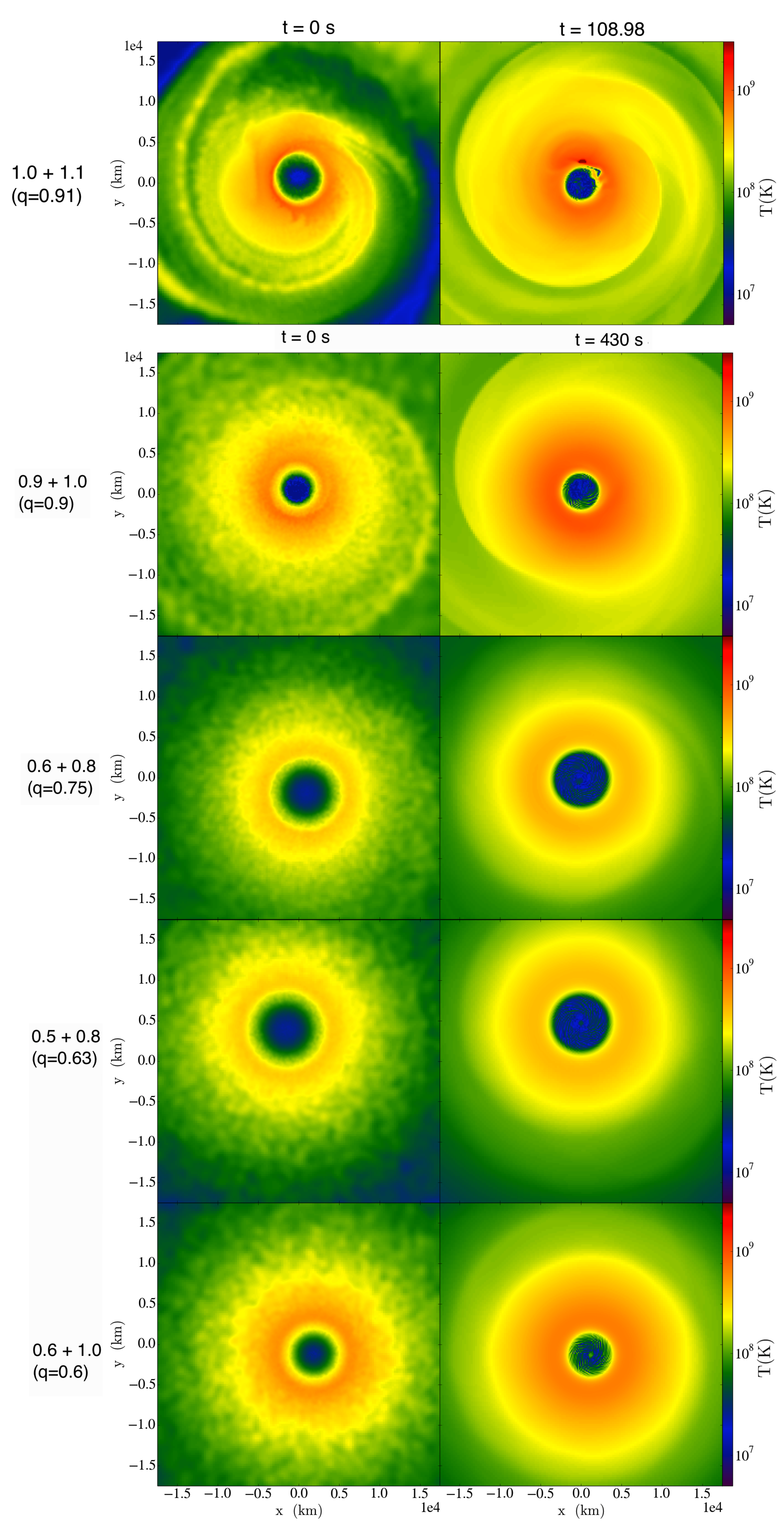}
		\caption{An expanded view of the initial and final disk temperatures of all models. Consistent with virial estimates, the maximum temperature is proportional to the primary WD mass and is inversely proportional to mass ratio, $q$. Also consistent with virial estimates, these plots show that the maximum temperature is generally at the boundary between the disk and the primary WD surface.}
		\label{fig:temp_disk}
	\end{center}
\end{figure}

\begin{figure}
	\begin{center}
		\includegraphics[width=1.0\columnwidth]{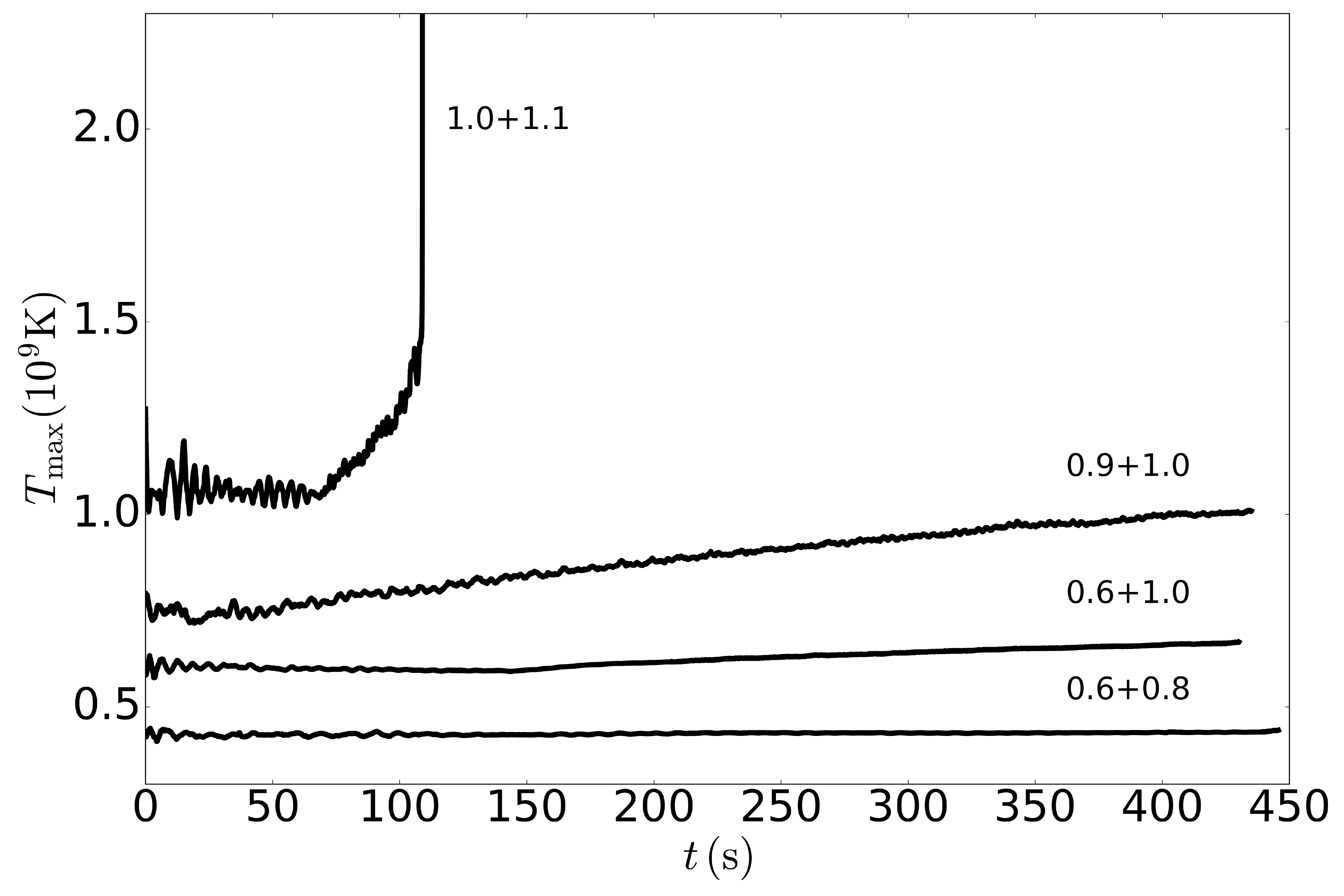}	
		\caption{ The evolution of the maximum temperature as a function of time for all models considered. The detonating model, $1.1 M_{\odot}$ + $1.0 M_{\odot}$, undergoes a rapid increase of maximum temperature, leading to a detonation at $t = 108.98$ s. A slower increase in temperature is also observed for the $0.9 M_{\odot}$ + $1.0 M_{\odot}$ model, although this does not lead to a detonation over the 430 s timescale considered. Here, systems with smaller total mass and mass-ratios do not show such increments. The  $0.5 M_{\odot}$ + $0.8 M_{\odot}$ model lies  close to  the evolution of the $0.6 M_{\odot}$ + $0.8 M_{\odot}$ model, and is not plotted for the sake of clarity.} 
		\label{fig:maxtemp_evol}
	\end{center}
\end{figure}
We develop a simple local runaway criteria by comparing the burning timescale ($t_{\rm burn}$) with the dynamical timescale ($t_{\rm dyn}$). Here we take the dynamical time $t_{\rm dyn} = (G \rho)^{-1/2}$ where $\rho$ is density of the cell. Following other authors \citep{Dan2014A}, we develop an analytic form for $t_{\rm burn} = c_{p}T/\dot{\epsilon}$ for carbon burning using a single-reaction rate. Here $\dot{\epsilon }$ is nuclear energy generation rate, $c_{p}$ is specific heat capacity at constant pressure and $T$ is the temperature.  The locus of these points satisfying  $t_{\rm burn} = t_{\rm dyn}$ yields the desired detonation condition as a curve in the density-temperature plane.  We plot the detonation criteria curve in Fig.~\ref{fig:det_region}, along with our simulation data. In the production of the analytical curves, it has been assumed that the nuclear reaction timescale is set by $^{12}$C ($^{12}$C,$\alpha$)  $^{20}$ N \citep{Blinnikov_1987A,Fowler_1988A}.
\begin{figure}
	\begin{center}
		\includegraphics[width=1.0\columnwidth]{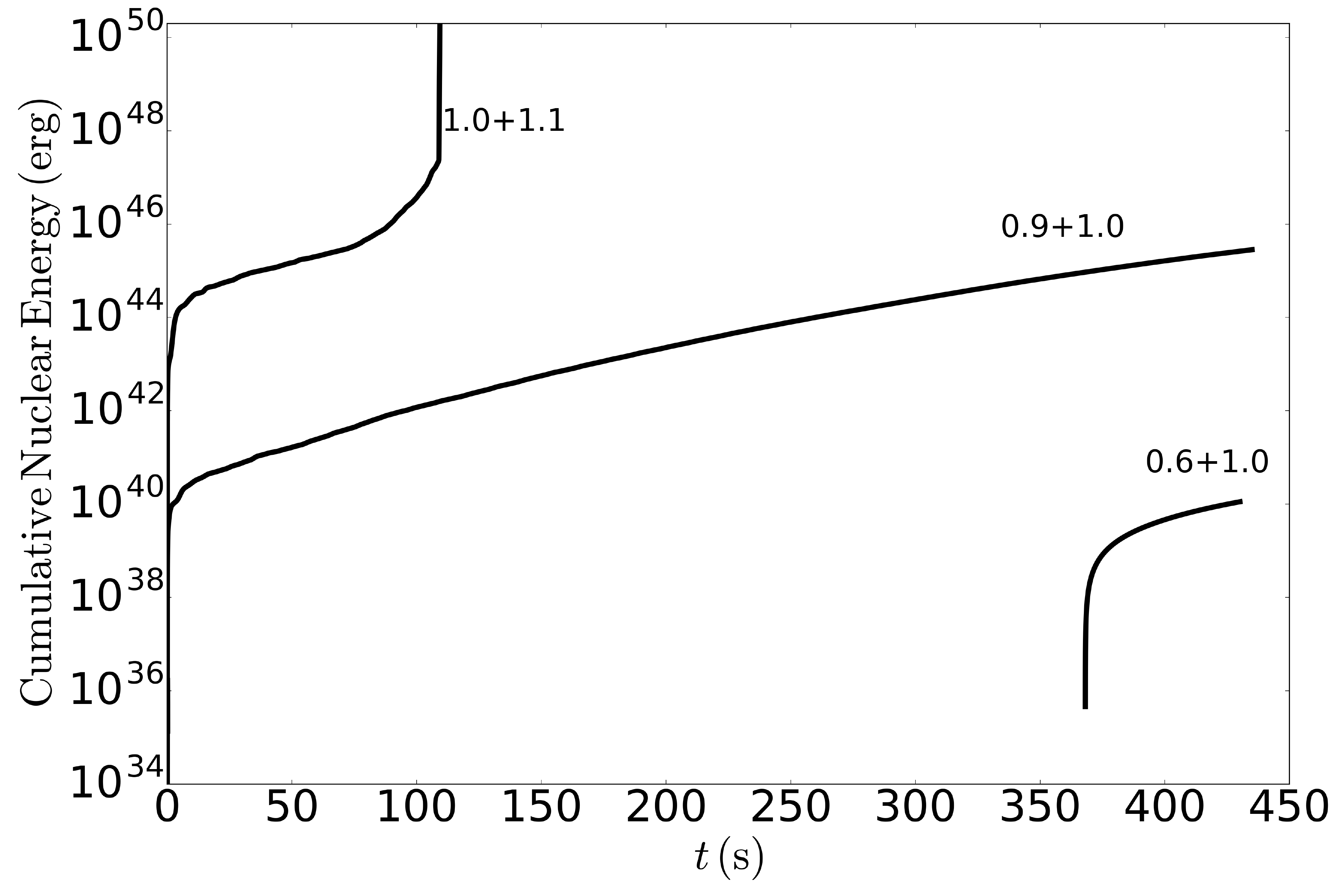}
		\caption{Evolution of the cumulative nuclear energy release. The models -- $1.1 M_{\odot}$ + $1.0 M_{\odot}$ , $0.9 M_{\odot}$ + $1.0 M_{\odot}$ and $0.5 M_{\odot}$ + $0.8 M_{\odot}$ have a positive value of the nuclear energy generation. Due to their lower temperatures, there is 
			a net cooling due to netrino losses in other models.} 
		\label{fig:energy_evol}
	\end{center}
\end{figure}

\begin{figure}
	\begin{center}
		\includegraphics[width=1.\columnwidth]{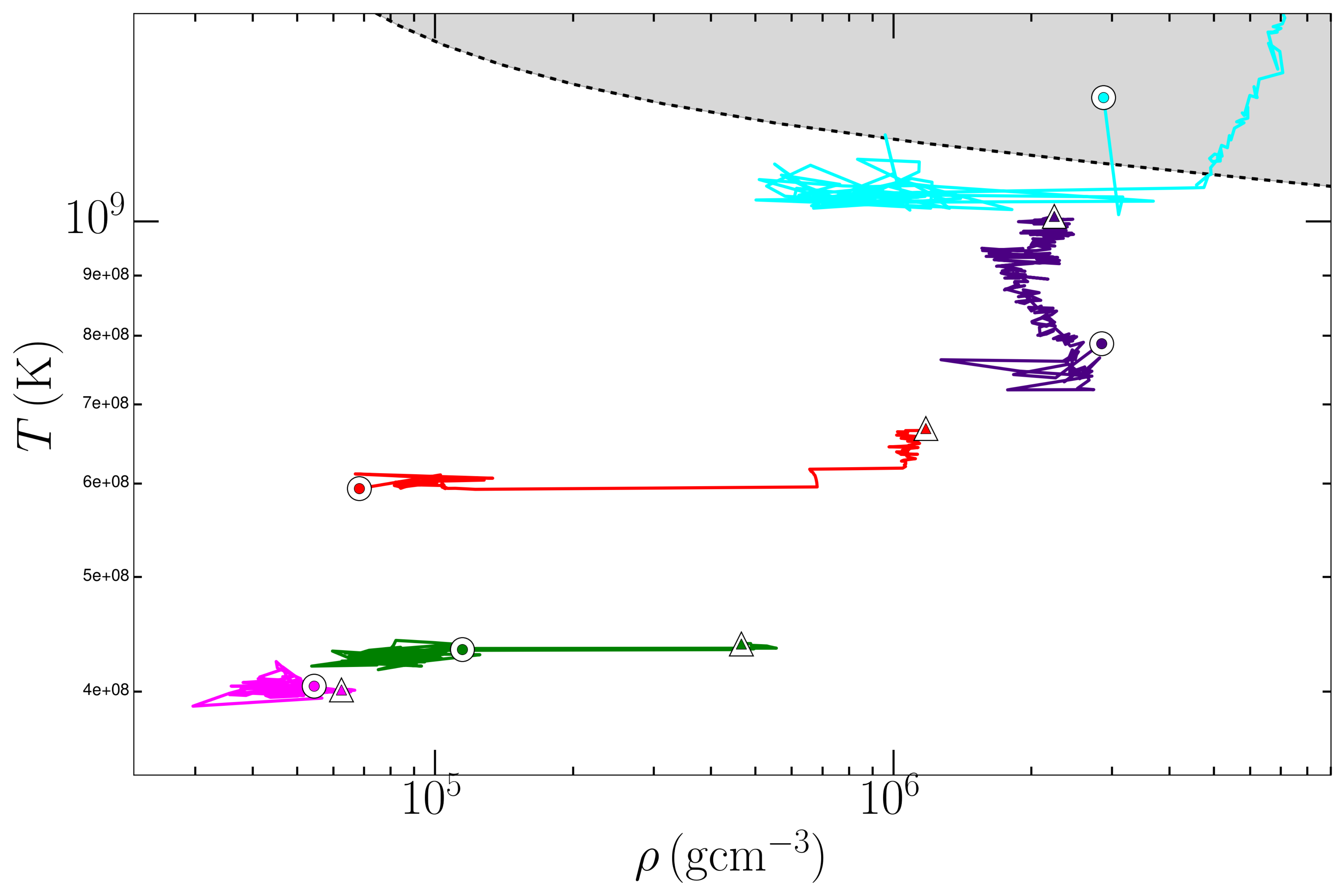}
		\caption{Evolution of the maximum temperature and the density on a log-log scale at the maximum temperature of four models we have simulated (cyan for $1.0 M_{\odot} + 1.1 M_{\odot}$, indigo for $0.9 M_{\odot} +  1.0 M_{\odot}$, red for $0.6 M_{\odot} +  1.0 M_{\odot}$, green for $0.6 M_{\odot} +  0.8 M_{\odot}$, magenta for $0.5 M_{\odot} +  0.8 M_{\odot}$). The curves are dense and noisy because the point of maximum temperature is not always the same, but corresponds to different (but nearby) cells. The empty circles and tringles represent the initial and final evolutionary points for the respective models. The dashed curve represents $t_{\rm dyn}/t_{\rm nuc}=1$ for carbon detonation, and the shaded region above it is the detonation regime as determined by this criterion. The detonating model  ($1.0 M_{\odot} + 1.1 M_{\odot}$), crosses carbon ignition region, while lower mass models secularly move towards the detonation criterion.  The final point of evolution for the model $1.0 M_{\odot} + 1.1 M_{\odot}$ is outside of the plot.}
		\label{fig:det_region}
	\end{center}
\end{figure}
As mentioned previously, in the models presented here, the maximum temperature is found to be just outside the surface of the primary WD (see Fig.~\ref{fig:temp_disk}). In the spiral mechanism, the hot material accretes at the surface of the white dwarf, mixing with the cold degenerate material and driving the temperature towards C ignition. Hence, we examine the temperature and density at the location of the maximum temperature in relation to the detonation criterion. We find that three lower mass models ($0.8\, M_{\sun} + 0.5 M_{\sun}$,$1.0\, M_{\sun} + 0.6 M_{\sun}$,$1.0\, M_{\sun} + 0.9 M_{\sun}$) are outside the C-ignition region (see Fig.~\ref{fig:det_region}). Because of their significant nuclear energy production, the $1.1\, M_{\sun} + 1.0 M_{\sun}$, $1.0\, M_{\sun} + 0.9 M_{\sun}$ and $1.0\, M_{\sun} + 0.6 M_{\sun}$ models generally experience an increase in their peak temperature. That is, these three models evolve towards higher temperature in the $(\rho, \rm{T}_{max})$ plane in figure \ref{fig:det_region}. We also observe that the region containing the hotspot lies inside the tidal radii of the mergers. We find mass accretion driven by the spiral modes accumulates mass near the surface of the white dwarf. As a result, the hotspots also move towards higher densities for models in figure \ref{fig:det_region}. The increase in density at peak temperature location is smaller for models with weaker spiral modes as shown for three of the lower mass models in figure \ref{fig:det_region}. However, two of the higher-mass models also have significant nuclear energy production which tends to lower their density at the location of peak temperature. These findings demonstrate the potential for the spiral mechanism to drive WD mergers towards  carbon ignition, and possible detonation, even when prompt detonation fails. Further studies are required to fully understand the long-term secular evolution in such post-merger sub-Chandrasekhar binary WD models.

%In these two models, this two effects tend to compete with each other which leads to detonation for $1.1\, M_{\sun} + 1.0 M_{\sun}$ case but, not for $1.0\, M_{\sun} + 0.9 M_{\sun}$ for the time of evolution considered here.
A preliminary estimation using maximum temperature and cumulative nuclear energy graph shows that the models $1.0\, M_{\sun} + 0.9 M_{\sun}$ and $1.0\, M_{\sun} + 0.6 M_{\sun}$ will reach the carbon detonation condition in 145 and 58 inner disk dynamical times respectively. We assume that the temperature-density condition for carbon detonation would be same for all of the high mass systems ($1.0 M_{\odot}+1.1 M_{\odot}, 0.9 M_{\odot}+1.0 M_{\odot}, 0.6 M_{\odot} +1.0 M_{\odot}$ ). 
%Moreover, carbon-detonation condition depends on density negligibly over the range ($2-4 \times 10^7$ g/cm$^3$) relevant in these three models. So, maximum temperature can be used to roughly estimate the detonation time.

%{\bf ($t_{dyn} = t_{burn}$)-curve of carbon burning (13 approx), pure He-burning (13 approx), He-burning with full reaction network (206-isotope)}
%We believe that double-detonation condition is most likely to be present during earlier phases of the merger for models presented in our paper. 

% discussion of which one would go ''.Ia'' and which one to SN Ia.

\section{Summary and Conclusions}
\label{conclusion}

We have examined the hydrodynamical evolution of binary white dwarf mergers with accretion disks formed from the disruption of the secondary WD. We find that binary WD systems undergoing unstable mass transfer with comparable masses generally develop a spiral mode instability subsequent to merger. \ignore{confirmations of our estimates of disk-stability to a great extent.} The accretion disk around the primary WD is subject to the eccentric gravitational instability, which leads to an $m=1$ spiral mode. We observe that the strength of all modes considered, from $m=1$ through $m = 4$, sensitively depends upon and grows with increasing mass ratios of the binary WD system.  We also have found that the temperature and density conditions do not reach carbon ignition condition over $\sim 8 - 22$ dynamical times for system masses less than $2.1\  M_{\sun}$. However, the slow secular increase of the maximum temperature in the model resulting from the merger of the $0.9\, M_{\sun} + 1.0\,  M_{\sun}$ and $1.0 + 0.6 M_{\odot}$  binary systems suggest that they could reach carbon detonation on longer timescales than explored in this study. \ignore{Nothing can be said about detonation in such models as many of our assumptions used in numerical simulations would not hold true at such time scales.} 

We have analyzed our results using an analytical criterion for runaway carbon burning. However, it has been shown using extended nuclear reaction network including additional $\alpha$-capture reaction pathways that the mass of the He layer required for detonation could be as small as $0.1\, M_{\sun}$ to $0.06\, M_{\sun}$ for WDs with masses $0.8\, M_{\sun}$ to $1.0\, M_{\sun}$, respectively \citep{Shen2014A}. Thus it may be the case that C/O WDs including such thin He layers, and modeled using an extended reaction network may result in qualitatively different outcomes than found here. The detonation of the He layer may lead to the detonation of the C/O core, or otherwise lead to a `.Ia'  SN. Future work is needed to determine the  outcome of these systems.

Our previous paper \citep{Kashyap2015A} showed that a model with a total system mass $2.1\, M_{\sun}$ produces $0.6\, M_{\sun}$ of $^{56}$Ni, which results in a normal, slowly-declining SN Ia similar to SN 2001ay \citep{vanrossumetal16}. In light of these previous findings, if lower-mass binary WD models are able to detonate, either through a slow secular onset as suggested in this paper, or through the ignition of thin He layers, they might span a range of luminosities and decline rates covering the full range of normal SNe Ia. Such double-degenerate mergers may provide a broader pathway for double-degenerate merger to contribute to the total SNe Ia rate, and help resolve outstanding issues regarding the trigger mechanism and the rates of double-degenerate SNe Ia.

{\bf Acknowledgements}  We thank James Guillochon, Daan Van Rossum, Chris Byrohl, and Pranav Dave for useful discussions. We also would like to thank the anonymous reviewer for their useful comments and insights. The work of EG-B, GA-S and PL-A was partially funded by MINECO AYA2014-59084-P grant and by the AGAUR. The software used in this work was in part developed by the DOE NNSA-ASC OASCR Flash Center at the University of Chicago. This work used the Extreme Science and Engineering discovery Environment (XSEDE), which is supported by National Science Foundation grant number ACI-1053575. Simulations at UMass Dartmouth were performed on a computer cluster supported by NSF grant CNS-0959382 and AFOSR DURIP grant FA9550-10-1-0354. RTF thanks the Institute for Theory and Computation at the Harvard-Smithsonian Center for Astrophysics, and the Kavli Institute for Theoretical Physics, supported in part by the national Science Foundation under grant NSF PHY11-25915, for visiting support during which this work was completed. This research has made use of resources from NASA's Astrophysics Data System and the yt astrophysics analysis software suite \citep{Turk_2011}.

%\bibliography{ref_SNIa,rtf_refs}

\begin{thebibliography}{}
	\expandafter\ifx\csname natexlab\endcsname\relax\def\natexlab#1{#1}\fi
	
	\bibitem[{Adams {et~al.}(1989)Adams, Ruden, \& Shu}]{Adams_1989}
	Adams, F.~C., Ruden, S.~P., \& Shu, F.~H. 1989, \apj, 347, 959
	
	\bibitem[{{Aznar-Sigu{\'a}n} {et~al.}(2013){Aznar-Sigu{\'a}n},
		{Garc{\'{\i}}a-Berro}, {Lor{\'e}n-Aguilar}, {Jos{\'e}}, \&
		{Isern}}]{Aznar_2013A}
	{Aznar-Sigu{\'a}n}, G., {Garc{\'{\i}}a-Berro}, E., {Lor{\'e}n-Aguilar}, P.,
	{Jos{\'e}}, J., \& {Isern}, J. 2013, \mnras, 434, 2539
	
	\bibitem[{{Aznar-Sigu{\'a}n} {et~al.}(2015){Aznar-Sigu{\'a}n},
		{Garc{\'{\i}}a-Berro}, {Lor{\'e}n-Aguilar}, {Soker}, \& {Kashi}}]{Aznar_2015}
	{Aznar-Sigu{\'a}n}, G., {Garc{\'{\i}}a-Berro}, E., {Lor{\'e}n-Aguilar}, P.,
	{Soker}, N., \& {Kashi}, A. 2015, \mnras, 450, 2948
	
	\bibitem[{{Badenes} \& {Maoz}(2012)}]{Badenes2012A}
	{Badenes}, C., \& {Maoz}, D. 2012, \apjl, 749, L11
	
	\bibitem[{{Blaes} \& {Hawley}(1988)}]{Blaes_1988}
	{Blaes}, O.~M., \& {Hawley}, J.~F. 1988, \apj, 326, 277
	
	\bibitem[{{Blinnikov} \& {Khokhlov}(1987)}]{Blinnikov_1987A}
	{Blinnikov}, S.~I., \& {Khokhlov}, A.~M. 1987, Pisma v Astronomicheskii
	Zhurnal, 13, 868
	
	\bibitem[{{Calder} {et~al.}(2002){Calder}, {Fryxell}, {Plewa}, {Rosner},
		{Dursi}, {Weirs}, {Dupont}, {Robey}, {Kane}, {Remington}, {Drake}, {Dimonte},
		{Zingale}, {Timmes}, {Olson}, {Ricker}, {MacNeice}, \& {Tufo}}]{Calder2002A}
	{Calder}, A.~C., {Fryxell}, B., {Plewa}, T., {et~al.} 2002, \apjs, 143, 201
	
	\bibitem[{{Caughlan} \& {Fowler}(1988)}]{Fowler_1988A}
	{Caughlan}, G.~R., \& {Fowler}, W.~A. 1988, Atomic Data and Nuclear Data
	Tables, 40, 283
	
	\bibitem[{{Childress} {et~al.}(2015){Childress}, {Hillier}, {Seitenzahl},
		{Sullivan}, {Maguire}, {Taubenberger}, {Scalzo}, {Ruiter}, {Blagorodnova},
		{Camacho}, {Castillo}, {Elias-Rosa}, {Fraser}, {Gal-Yam}, {Graham}, {Howell},
		{Inserra}, {Jha}, {Kumar}, {Mazzali}, {McCully}, {Morales-Garoffolo},
		{Pandya}, {Polshaw}, {Schmidt}, {Smartt}, {Smith}, {Sollerman}, {Spyromilio},
		{Tucker}, {Valenti}, {Walton}, {Wolf}, {Yaron}, {Young}, {Yuan}, \&
		{Zhang}}]{Childress_2015A}
	{Childress}, M.~J., {Hillier}, D.~J., {Seitenzahl}, I., {et~al.} 2015, ArXiv
	e-prints, arXiv:1507.02501
	
	\bibitem[{{Couch} {et~al.}(2013){Couch}, {Graziani}, \& {Flocke}}]{Couch_2013}
	{Couch}, S.~M., {Graziani}, C., \& {Flocke}, N. 2013, \apj, 778, 181
	
	\bibitem[{{Cyburt} {et~al.}(2010){Cyburt}, {Amthor}, {Ferguson}, {Meisel},
		{Smith}, {Warren}, {Heger}, {Hoffman}, {Rauscher}, {Sakharuk}, {Schatz},
		{Thielemann}, \& {Wiescher}}]{Cyburt_2010}
	{Cyburt}, R.~H., {Amthor}, A.~M., {Ferguson}, R., {et~al.} 2010, \apjs, 189,
	240
	
	\bibitem[{{Dan} {et~al.}(2014){Dan}, {Rosswog}, {Br{\"u}ggen}, \&
		{Podsiadlowski}}]{Dan2014A}
	{Dan}, M., {Rosswog}, S., {Br{\"u}ggen}, M., \& {Podsiadlowski}, P. 2014,
	\mnras, 438, 14
	
	\bibitem[{{Dan} {et~al.}(2011){Dan}, {Rosswog}, {Guillochon}, \&
		{Ramirez-Ruiz}}]{Dan2011A}
	{Dan}, M., {Rosswog}, S., {Guillochon}, J., \& {Ramirez-Ruiz}, E. 2011, \apj,
	737, 89
	
	\bibitem[{Dubey {et~al.}(2009)Dubey, Antypas, Ganapathy, Reid, Riley, Sheeler,
		Siegel, \& Weide}]{Dubey_2009A}
	Dubey, A., Antypas, K., Ganapathy, M.~K., {et~al.} 2009, Parallel Computing,
	35, 512
	
	\bibitem[{{Fisher} \& {Jumper}(2015)}]{Fisher2015A}
	{Fisher}, R., \& {Jumper}, K. 2015, \apj, 805, 150
	
	\bibitem[{{Fryxell} {et~al.}(2000){Fryxell}, {Olson}, {Ricker}, {Timmes},
		{Zingale}, {Lamb}, {MacNeice}, {Rosner}, {Truran}, \& {Tufo}}]{Fryxell2000A}
	{Fryxell}, B., {Olson}, K., {Ricker}, P., {et~al.} 2000, \apjs, 131, 273
	
	\bibitem[{{Guerrero} {et~al.}(2004){Guerrero}, {Garc{\'{\i}}a-Berro}, \&
		{Isern}}]{Guerrero_2004A}
	{Guerrero}, J., {Garc{\'{\i}}a-Berro}, E., \& {Isern}, J. 2004, \aap, 413, 257
	
	\bibitem[{{Hahn} {et~al.}(2010){Hahn}, {Teyssier}, \& {Carollo}}]{hahnetal10}
	{Hahn}, O., {Teyssier}, R., \& {Carollo}, C.~M. 2010, \mnras, 405, 274
	
	\bibitem[{{Hix} {et~al.}(1998){Hix}, {Khokhlov}, {Wheeler}, \&
		{Thielemann}}]{Hix_1998A}
	{Hix}, W.~R., {Khokhlov}, A.~M., {Wheeler}, J.~C., \& {Thielemann}, F.-K. 1998,
	\apj, 503, 332
	
	\bibitem[{{Iben} \& {Tutukov}(1984)}]{Iben_Tutukov_1984}
	{Iben}, Jr., I., \& {Tutukov}, A.~V. 1984, \apjs, 54, 335
	
	\bibitem[{{Ilkov} \& {Soker}(2013)}]{Ilkov_2013}
	{Ilkov}, M., \& {Soker}, N. 2013, \mnras, 428, 579
	
	\bibitem[{{Itoh} {et~al.}(1996){Itoh}, {Hayashi}, {Nishikawa}, \&
		{Kohyama}}]{Itoh_1996}
	{Itoh}, N., {Hayashi}, H., {Nishikawa}, A., \& {Kohyama}, Y. 1996, \apjs, 102,
	411
	
	\bibitem[{{Ju} {et~al.}(2016){Ju}, {Stone}, \& {Zhu}}]{Ju2016A}
	{Ju}, W., {Stone}, J.~M., \& {Zhu}, Z. 2016, \apj, 823, 81
	
	\bibitem[{{Kashi} \& {Soker}(2011)}]{Kashi_2011}
	{Kashi}, A., \& {Soker}, N. 2011, \mnras, 417, 1466
	
	\bibitem[{{Kashyap} {et~al.}(2015){Kashyap}, {Fisher}, {Garc{\'{\i}}a-Berro},
		{Aznar-Sigu{\'a}n}, {Ji}, \& {Lor{\'e}n-Aguilar}}]{Kashyap2015A}
	{Kashyap}, R., {Fisher}, R., {Garc{\'{\i}}a-Berro}, E., {et~al.} 2015, \apjl,
	800, L7
	
	\bibitem[{{Krumholz} {et~al.}(2004){Krumholz}, {McKee}, \&
		{Klein}}]{Krumholz2004}
	{Krumholz}, M.~R., {McKee}, C.~F., \& {Klein}, R.~I. 2004, \apj, 611, 399
	
	\bibitem[{{Kushnir} {et~al.}(2013){Kushnir}, {Katz}, {Dong}, {Livne}, \&
		{Fern{\'a}ndez}}]{Kushnir2013A}
	{Kushnir}, D., {Katz}, B., {Dong}, S., {Livne}, E., \& {Fern{\'a}ndez}, R.
	2013, \apjl, 778, L37
	
	\bibitem[{{Laughlin} {et~al.}(1997){Laughlin}, {Korchagin}, \&
		{Adams}}]{Laughlin1997}
	{Laughlin}, G., {Korchagin}, V., \& {Adams}, F.~C. 1997, \apj, 477, 410
	
	\bibitem[{{Livio} \& {Riess}(2003)}]{Livio_2003}
	{Livio}, M., \& {Riess}, A.~G. 2003, \apjl, 594, L93
	
	\bibitem[{{Livne}(1990)}]{Livne1990A}
	{Livne}, E. 1990, \apjl, 354, L53
	
	\bibitem[{{Livne} \& {Arnett}(1995)}]{Livne1995A}
	{Livne}, E., \& {Arnett}, D. 1995, \apj, 452, 62
	
	\bibitem[{{Lor{\'e}n-Aguilar} {et~al.}(2009){Lor{\'e}n-Aguilar}, {Isern}, \&
		{Garc{\'{\i}}a-Berro}}]{Loren2009A}
	{Lor{\'e}n-Aguilar}, P., {Isern}, J., \& {Garc{\'{\i}}a-Berro}, E. 2009, \aap,
	500, 1193
	
	\bibitem[{Lor{\'{e}}n-Aguilar {et~al.}(2010)Lor{\'{e}}n-Aguilar, Isern, \&
		Garc{\'{\i}}a-Berro}]{Loren_Aguilar_2010}
	Lor{\'{e}}n-Aguilar, P., Isern, J., \& Garc{\'{\i}}a-Berro, E. 2010, \mnras,
	406, 2749{\textendash}2763
	
	\bibitem[{{MacLeod} {et~al.}(2015){MacLeod}, {Guillochon}, {Ramirez-Ruiz},
		{Kasen}, \& {Rosswog}}]{Guillochon_2015A}
	{MacLeod}, M., {Guillochon}, J., {Ramirez-Ruiz}, E., {Kasen}, D., \& {Rosswog},
	S. 2015, ArXiv e-prints, arXiv:1508.02399
	
	\bibitem[{{Maoz} \& {Badenes}(2010)}]{maozbadenes_2010}
	{Maoz}, D., \& {Badenes}, C. 2010, \mnras, 407, 1314
	
	\bibitem[{{Maoz} {et~al.}(2013){Maoz}, {Mannucci}, \&
		{Nelemans}}]{Maozetal_2013}
	{Maoz}, D., {Mannucci}, F., \& {Nelemans}, G. 2013, ArXiv e-prints
	
	\bibitem[{{Miyaji} {et~al.}(1980){Miyaji}, {Nomoto}, {Yokoi}, \&
		{Sugimoto}}]{Miyaji1980A}
	{Miyaji}, S., {Nomoto}, K., {Yokoi}, K., \& {Sugimoto}, D. 1980, \pasj, 32, 303
	
	\bibitem[{{Moll} {et~al.}(2014){Moll}, {Raskin}, {Kasen}, \&
		{Woosley}}]{molletal14}
	{Moll}, R., {Raskin}, C., {Kasen}, D., \& {Woosley}, S.~E. 2014, \apj, 785, 105
	
	\bibitem[{{Nelemans} {et~al.}(2001){Nelemans}, {Yungelson}, {Portegies Zwart},
		\& {Verbunt}}]{Nelemans2001A}
	{Nelemans}, G., {Yungelson}, L.~R., {Portegies Zwart}, S.~F., \& {Verbunt}, F.
	2001, \aap, 365, 491
	
	\bibitem[{{Nelson} {et~al.}(1998){Nelson}, {Benz}, {Adams}, \&
		{Arnett}}]{Nelson1998}
	{Nelson}, A.~F., {Benz}, W., {Adams}, F.~C., \& {Arnett}, D. 1998, \apj, 502,
	342
	
	\bibitem[{{Pakmor} {et~al.}(2010){Pakmor}, {Kromer}, {R{\"o}pke}, {Sim},
		{Ruiter}, \& {Hillebrandt}}]{Pakmor_2010}
	{Pakmor}, R., {Kromer}, M., {R{\"o}pke}, F.~K., {et~al.} 2010, \nat, 463, 61
	
	\bibitem[{{Pakmor} {et~al.}(2012{\natexlab{a}}){Pakmor}, {Kromer},
		{Taubenberger}, {Sim}, {R{\"o}pke}, \& {Hillebrandt}}]{Pakmor2012A}
	{Pakmor}, R., {Kromer}, M., {Taubenberger}, S., {et~al.} 2012{\natexlab{a}},
	\apjl, 747, L10
	
	\bibitem[{{Pakmor} {et~al.}(2012{\natexlab{b}}){Pakmor}, {Kromer},
		{Taubenberger}, {Sim}, {R{\"o}pke}, \& {Hillebrandt}}]{Pakmor_2012}
	---. 2012{\natexlab{b}}, \apjl, 747, L10
	
	\bibitem[{{Pakmor} {et~al.}(2013){Pakmor}, {Kromer}, {Taubenberger}, \&
		{Springel}}]{Pakmor2013A}
	{Pakmor}, R., {Kromer}, M., {Taubenberger}, S., \& {Springel}, V. 2013, \apjl,
	770, L8
	
	\bibitem[{{Perlmutter} {et~al.}(1999){Perlmutter}, {Aldering}, {Goldhaber},
		{Knop}, {Nugent}, {Castro}, {Deustua}, {Fabbro}, {Goobar}, {Groom}, {Hook},
		{Kim}, {Kim}, {Lee}, {Nunes}, {Pain}, {Pennypacker}, {Quimby}, {Lidman},
		{Ellis}, {Irwin}, {McMahon}, {Ruiz-Lapuente}, {Walton}, {Schaefer}, {Boyle},
		{Filippenko}, {Matheson}, {Fruchter}, {Panagia}, {Newberg}, {Couch}, \&
		{Supernova Cosmology Project}}]{Perlmutter1999A}
	{Perlmutter}, S., {Aldering}, G., {Goldhaber}, G., {et~al.} 1999, \apj, 517,
	565
	
	\bibitem[{{Raskin} {et~al.}(2012{\natexlab{a}}){Raskin}, {Scannapieco},
		{Fryer}, {Rockefeller}, \& {Timmes}}]{Raskin2012A}
	{Raskin}, C., {Scannapieco}, E., {Fryer}, C., {Rockefeller}, G., \& {Timmes},
	F.~X. 2012{\natexlab{a}}, \apj, 746, 62
	
	\bibitem[{{Raskin} {et~al.}(2012{\natexlab{b}}){Raskin}, {Scannapieco},
		{Fryer}, {Rockefeller}, \& {Timmes}}]{raskinetal12}
	---. 2012{\natexlab{b}}, \apj, 746, 62
	
	\bibitem[{{Raskin} {et~al.}(2009){Raskin}, {Timmes}, {Scannapieco}, {Diehl}, \&
		{Fryer}}]{Raskin2009A}
	{Raskin}, C., {Timmes}, F.~X., {Scannapieco}, E., {Diehl}, S., \& {Fryer}, C.
	2009, \mnras, 399, L156
	
	\bibitem[{{Riess} {et~al.}(1998){Riess}, {Filippenko}, {Challis},
		{Clocchiatti}, {Diercks}, {Garnavich}, {Gilliland}, {Hogan}, {Jha},
		{Kirshner}, {Leibundgut}, {Phillips}, {Reiss}, {Schmidt}, {Schommer},
		{Smith}, {Spyromilio}, {Stubbs}, {Suntzeff}, \& {Tonry}}]{Riess1998A}
	{Riess}, A.~G., {Filippenko}, A.~V., {Challis}, P., {et~al.} 1998, \aj, 116,
	1009
	
	\bibitem[{{Rosswog} {et~al.}(2009){Rosswog}, {Kasen}, {Guillochon}, \&
		{Ramirez-Ruiz}}]{Rosswog2009A}
	{Rosswog}, S., {Kasen}, D., {Guillochon}, J., \& {Ramirez-Ruiz}, E. 2009,
	\apjl, 705, L128
	
	\bibitem[{{Rueda} {et~al.}(2013){Rueda}, {Boshkayev}, {Izzo}, {Ruffini},
		{Lor{\'e}n-Aguilar}, {K{\"u}lebi}, {Aznar-Sigu{\'a}n}, \&
		{Garc{\'{\i}}a-Berro}}]{Rueda_2013A}
	{Rueda}, J.~A., {Boshkayev}, K., {Izzo}, L., {et~al.} 2013, \apjl, 772, L24
	
	\bibitem[{{Saio} \& {Nomoto}(1985)}]{Saio1985A}
	{Saio}, H., \& {Nomoto}, K. 1985, \aap, 150, L21
	
	\bibitem[{{Schwab} {et~al.}(2012){Schwab}, {Shen}, {Quataert}, {Dan}, \&
		{Rosswog}}]{Schwab2012A}
	{Schwab}, J., {Shen}, K.~J., {Quataert}, E., {Dan}, M., \& {Rosswog}, S. 2012,
	\mnras, 427, 190
	
	\bibitem[{{Seitenzahl} {et~al.}(2009){Seitenzahl}, {Meakin}, {Townsley},
		{Lamb}, \& {Truran}}]{Seitenzahl_2009}
	{Seitenzahl}, I.~R., {Meakin}, C.~A., {Townsley}, D.~M., {Lamb}, D.~Q., \&
	{Truran}, J.~W. 2009, \apj, 696, 515
	
	\bibitem[{{Shen} {et~al.}(2012){Shen}, {Bildsten}, {Kasen}, \&
		{Quataert}}]{Shen2012A}
	{Shen}, K.~J., {Bildsten}, L., {Kasen}, D., \& {Quataert}, E. 2012, \apj, 748,
	35
	
	\bibitem[{{Shen} \& {Moore}(2014)}]{Shen2014A}
	{Shen}, K.~J., \& {Moore}, K. 2014, \apj, 797, 46
	
	\bibitem[{Shu {et~al.}(1990)Shu, Tremaine, Adams, \& Ruden}]{Shu_1990}
	Shu, F.~H., Tremaine, S., Adams, F.~C., \& Ruden, S.~P. 1990, \apj, 358, 495
	
	\bibitem[{{Sparks} \& {Stecher}(1974)}]{Sparks_1974}
	{Sparks}, W.~M., \& {Stecher}, T.~P. 1974, \apj, 188, 149
	
	\bibitem[{{Springel}(2010)}]{springel10}
	{Springel}, V. 2010, \araa, 48, 391
	
	\bibitem[{{Tasker} {et~al.}(2008){Tasker}, {Brunino}, {Mitchell}, {Michielsen},
		{Hopton}, {Pearce}, {Bryan}, \& {Theuns}}]{taskeretal08}
	{Tasker}, E.~J., {Brunino}, R., {Mitchell}, N.~L., {et~al.} 2008, \mnras, 390,
	1267
	
	\bibitem[{{Thompson}(2011)}]{Thompson2011A}
	{Thompson}, T.~A. 2011, \apj, 741, 82
	
	\bibitem[{Timmes(1999)}]{Timmes1999A}
	Timmes, F.~X. 1999, \apjs, 124, 241{\textendash}263
	
	\bibitem[{{Timmes} \& {Swesty}(2000)}]{Timmes2000A}
	{Timmes}, F.~X., \& {Swesty}, F.~D. 2000, \apjs, 126, 501
	
	\bibitem[{{Toonen} {et~al.}(2012){Toonen}, {Nelemans}, \& {Portegies
			Zwart}}]{Toonen_2012A}
	{Toonen}, S., {Nelemans}, G., \& {Portegies Zwart}, S. 2012, \aap, 546, A70
	
	\bibitem[{{Truelove} {et~al.}(1998){Truelove}, {Klein}, {McKee}, {Holliman},
		{Howell}, {Greenough}, \& {Woods}}]{trueloveetal98}
	{Truelove}, J.~K., {Klein}, R.~I., {McKee}, C.~F., {et~al.} 1998, \apj, 495,
	821
	
	\bibitem[{{Turk} {et~al.}(2011){Turk}, {Smith}, {Oishi}, {Skory}, {Skillman},
		{Abel}, \& {Norman}}]{Turk_2011}
	{Turk}, M.~J., {Smith}, B.~D., {Oishi}, J.~S., {et~al.} 2011, \apjs, 192, 9
	
	\bibitem[{{van Rossum} {et~al.}(2016){van Rossum}, {Kashyap}, {Fisher},
		{Wollaeger}, {Garc{\'{\i}}a-Berro}, {Aznar-Sigu{\'a}n}, {Ji}, \&
		{Lor{\'e}n-Aguilar}}]{vanrossumetal16}
	{van Rossum}, D.~R., {Kashyap}, R., {Fisher}, R., {et~al.} 2016, \apj, 827, 128
	
	\bibitem[{{Webbink}(1984)}]{Webbink_1984A}
	{Webbink}, R.~F. 1984, \apj, 277, 355
	
	\bibitem[{{Whelan} \& {Iben}(1973)}]{Whelan_Iben_1973}
	{Whelan}, J., \& {Iben}, Jr., I. 1973, \apj, 186, 1007
	
	\bibitem[{{Woosley} \& {Weaver}(1994)}]{Woosley1994A}
	{Woosley}, S.~E., \& {Weaver}, T.~A. 1994, \apj, 423, 371
	
	\bibitem[{{Yoon} {et~al.}(2007){Yoon}, {Podsiadlowski}, \&
		{Rosswog}}]{Yoon2007A}
	{Yoon}, S.-C., {Podsiadlowski}, P., \& {Rosswog}, S. 2007, \mnras, 380, 933
	
	\bibitem[{{Zel'dovich} {et~al.}(1970){Zel'dovich}, {Librovich}, {Makhviladze},
		\& {Sivashinskil}}]{Zeldovich_1970A}
	{Zel'dovich}, Y.~B., {Librovich}, V.~B., {Makhviladze}, G.~M., \&
	{Sivashinskil}, G.~I. 1970, Journal of Applied Mechanics and Technical
	Physics, 11, 264
	
\end{thebibliography}

\end{document}